\documentstyle[twocolumn,aps,prl]{revtex}
\begin{document}
\draft
\title{Comment on ``Non-Contextual Hidden Variables and
Physical Measurements''\thanks{Submitted to Phys. Rev. Lett.}}
\author{Ad\'{a}n Cabello\thanks{Electronic address:
fite1z1@sis.ucm.es}}
\address{Departamento de F\'{\i}sica Aplicada,
Universidad de Sevilla, 41012 Sevilla, Spain}
\date{\today}
\maketitle
\begin{abstract}
Kent's conclusion that ``non-contextual hidden variable theories
cannot be excluded by theoretical arguments of
the Kochen-Specker type
once the imprecision in real world experiments is taken
into account'' [Phys. Rev. Lett. {\bf 83}, 3755 (1999)],
is criticized.
The Kochen-Specker theorem just points out that
it is impossible
even {\em to conceive} a hidden variable model in which the
outcomes of {\em all}
measurements are pre-determined; it does not matter if
these measurements are performed or not, or even if
these measurements can be achieved only with finite precision.
\end{abstract}

\narrowtext
In a recent Letter \cite{Kent99}, Kent generalizes a result advanced by
Meyer \cite{Meyer99}, and concludes that:
\mbox{(i) ``Non-contextual} hidden variable [NCHV] theories
cannot be excluded by theoretical arguments of
the \mbox{K[ochen-]S[pecker]} type \cite{Specker60,Bell66,KS67}
once the imprecision in real world experiments is taken
into account''.
(ii) ``This does not (...) affect the situation regarding local hidden
variable [LHV] theories, which can be refuted by experiment,
modulo reasonable assumptions \cite{Bell64,CHSH69,Aspect81}.''

In my view, the situation is the opposite:
The KS theorem
holds, precisely because it is a theoretical argument which deals
with {\em gedanken} concepts such as
ideal yes-no questions. However, the
empirical refutation of LHV theories
can be questioned precisely on the grounds of the inevitable finiteness of
the precision of real measurements. Allow me to illustrate both points.

The KS theorem is a mathematical statement which
asserts that for a physical
system described in quantum mechanics (QM) by a
Hilbert space of dimension greater than or equal to three,
it is possible to find a set of $n$ projection operators,
which represent yes-no questions about an individual physical system,
so that none of the $2^n$ possible sets of ``yes'' or ``no'' answers
is compatible with the sum rule of QM for orthogonal resolutions of
the identity (i.e., if the sum of a subset of mutually orthogonal
projection operators is the identity, one and only one of the
corresponding answers ought to be ``yes'')  \cite{Peres93}.
The smallest example currently known
of such a set has only $18$ yes-no questions
(about a physical system described by a four-dimensional
Hilbert space) \cite{CEG96}.
As far as I can see, the plain new result contained in \cite{Kent99}
is the following: For any physical system described by a
finite Hilbert space, it is always possible
to construct a set of projection operators, which is {\em dense} in
the set of all projection operators,
so that an assignation of ``yes'' or ``no'' answers is possible
in a way compatible with the sum rule of QM.
From a mathematical point of view, it is clear that this
new result does not, by no means, nullify the KS theorem.
However, Kent affirms that this is so
when one takes into account that realistic physical measurements
are always of finite precision.
Kent seems to assume that the KS theorem concerns the results of a
(counterfactual) set of measurements, instead of
(the plain non-existence of) a set of yes-no questions with
pre-determined answers.
The KS theorem just points out that
it is impossible
even {\em to conceive} a hidden variable model in which the
outcomes of {\em all}
measurements are pre-determined; it does not matter if
these measurements are performed or not, or even if
these measurements can be achieved only with finite precision.
The KS theorem assumes that
any NCHV theory is a classical theory,
and since in classical physics
there is in principle no difficulty to conceive
ideal (i.e., defined with infinite precision) yes-no questions, then
it is quite legitimate to handle ideal yes-no questions
when one is trying to prove that
such a theory does not even exist.

The only possible loophole in the KS theorem would be
caused by the nonexistence of
some of the yes-no questions involved in
any of its proofs.
However, this loophole would have
very weird consequences. For instance, consider
a physical system described by a
three-dimensional real Hilbert space,
and assume that the only yes-no questions with a
real existence would be those
represented by projection operators defined
by vectors with {\em rational} components.
This subset is dense in the set of yes-no questions and
admits an assignation of ``yes'' or
``no'' answers compatible with the sum rule \cite{Meyer99}.
However, the initial assumption is
in conflict with the superposition principle because
some linear combinations of ``legal'' yes-no questions
would be illegal, since
their normalization
would demand irrational components \cite{Peresc}.

On the other hand,
the finite precision measurement problem matters in real experiments.
It will affect any real
experiment based on the KS theorem \cite{CG98}, and
indeed affects the theoretical
analysis of any real experiment to refute LHV.
In fact, real experiments like
those of Aspect {\em et al.} mentioned by Kent,
{\em admit} LHV models \cite{Santos91}.
These models still work even assuming
perfect efficiency of detectors, but vanish when
infinite precision of preparations and (of all required) measurements
is assumed.

The author thanks A. Peres and E. Santos
for useful comments and clarifications.
This work was supported
by the Universidad de Sevilla (Grant No.\ OGICYT-191-97)
and the Junta de Andaluc\'{\i}a (Grant No.\ FQM-239).


\begin{thebibliography}{99}
\bibitem{Kent99} A. Kent,
Phys. Rev. Lett. {\bf 83}, 3755 (1999);
Los Alamos e-print archive,
quant-ph/9906006.
\bibitem{Meyer99} D. A. Meyer,
Phys. Rev. Lett. {\bf 83}, 3751 (1999);
Los Alamos e-print archive,
quant-ph/9905080.
\bibitem{Specker60} E. P. Specker,
Dialectica {\bf 14}, 239 (1960).
English version in {\em The Logico-Algebraic Approach to
Quantum Mechanics. Volume I: Historical Evolution},
edited by C. A. Hooker
(Reidel, Dordrecht, 1975), p. 135.
\bibitem{Bell66} J. S. Bell,
Rev. Mod. Phys. {\bf 38}, 447 (1966).
\bibitem{KS67} S. Kochen and E. P. Specker,
J. Math. Mech. {\bf 17}, 59 (1967).
Reprinted in {\em The Logico-Algebraic Approach to
Quantum Mechanics. Volume I: Historical Evolution},
edited by C. A. Hooker
(Reidel, Dordrecht, 1975), p. 293.
\bibitem{Bell64} J. S. Bell,
Physics (Long Island City, N.Y.) {\bf 1}, 195 (1964).
\bibitem{CHSH69} J. F. Clauser, M. A. Horne, A. Shimony, and R. A.
Holt, Phys. Rev. Lett. {\bf 23}, 880 (1969).
\bibitem{Aspect81} A. Aspect, P. Grangier, and G. Roger,
Phys. Rev. Lett. {\bf 47}, 460 (1981).
\bibitem{Peres93} A. Peres,
{\em Quantum Theory: Concepts and Methods}
(Kluwer, Dordrecht, 1993).
\bibitem{CEG96} A. Cabello,
J. M. Estebaranz, and G. Garc\'{\i}a Alcaine,
Phys. Lett. A {\bf 212}, 183 (1996).
\bibitem{Peresc} A. Peres, private communication.
\bibitem{CG98} A. Cabello and G. Garc\'{\i}a Alcaine,
Phys. Rev. Lett. {\bf 80}, 1797 (1998).
\bibitem{Santos91} E. Santos,
Phys. Rev. Lett. {\bf 66}, 1388 (1991);
Phys. Rev. Lett. {\bf 68}, 2702 (1992);
Phys. Rev. A {\bf 46}, 3646 (1992).

\end{thebibliography}
\end{document}